\documentclass{PoS}
\usepackage{caption}
\usepackage{subcaption}
\usepackage{url}
\DeclareMathAlphabet{\mathcal}{OMS}{cmsy}{m}{n}
\newcommand{\doublefig}[9]{
	\begin{figure*}[h!]
		\begin{subfigure}[t]{0.49\textwidth}
			\includegraphics[width=\textwidth,height=#1\textwidth]{#2}
			\caption{{\small #3.}}
			\label{#4}
		\end{subfigure}
		\begin{subfigure}[t]{0.49\textwidth}
			\includegraphics[width=\textwidth,height=#1\textwidth]{#5}
			\caption{{\small #6.}}
			\label{#7}
		\end{subfigure}
		\caption{{\small #8.}}
		\label{#9}
	\end{figure*}
}
\newcommand{\onefig}[5]{
	\begin{figure}[h!]
		\centering
		\includegraphics[width=#1\textwidth,height=#2\textwidth]{#3}
		\caption{{\small #4.}}
		\label{#5}
	\end{figure}
}
\newcommand{\esqtab}[6]{
	\begin{table}[h]
		\begin{center}
			\scalebox{#3}[#4]{
				\begin{tabular}{#5}
					#6
				\end{tabular}
			}	
			\caption{{\small #1.}}
			\label{#2}
		\end{center}
	\end{table}
}
\title{New gamma/hadron separation parameters for a neural network for HAWC}
\ShortTitle{New G/H separation parameter for a NN for HAWC}

\author{E. Bourbeau$^a$ ,\speaker{T. Capistr\'an}$^b$,I. Torres$^b$, E. Moreno$^c$ for the HAWC Collaboration$^d$ \\
        \llap{$^a$}McGill University, 45 Sherbrooke Street West, Montreal, Quebec, Canada \\
        \llap{$^b$}Instituto Nacional de Astrof\'isica, \'Optica y Electr\'onica, Luis Enrique Erro 1, Tonantzintla, Puebla 72840, M\'exico \\
	\llap{$^c$}Facultad de Ciencias F\'isico Matem\'aticas, Benem\'erita Universidad Aut\'onoma de Puebla, Ciudad Universitaria, Colonia San Manuel Puebla, M\'exico \\
        \llap{$^d$}For a complete author list, see \href{http://www.hawc-observatory.org/collaboration/icrc2017.php}{http://www.hawc-observatory.org/collaboration/icrc2017.php}.\\
        Email: \email{etienne.bourbeau@gmail.com}, \email{tcapistran@inaoep.mx}, \email{ibrahim@inaoep.mx}, \email{emoreno@fcfm.buap.mx}}

\abstract{The High-Altitude Water Cherenkov experiment (HAWC) observatory is located 4100 meters above sea level. HAWC is able to detect secondary particles from extensive air showers (EAS) initiated in the interaction of a primary particle (either a gamma or a charged cosmic ray) with the upper atmosphere. Because an overwhelming majority of EAS events are triggered by cosmic rays, background noise suppression plays an important role in the data analysis process of the HAWC observatory. Currently, HAWC uses cuts on two parameters (whose values depend on the spatial distribution and luminosity of an event) to separate gamma-ray events from background hadronic showers. In this work, a search for additional gamma-hadron separation parameters was conducted to improve the efficiency of the HAWC background suppression technique. The best-performing parameters were integrated to a feed-foward Multilayer Perceptron Neural Network (MLP-NN), along with the traditional parameters. Various iterations of MLP-NN's were trained on Monte Carlo data, and tested on Crab data. Preliminary results show that the addition of new parameters can improve the significance of the point source at high-energies (\textasciitilde{} TeV), at the expense of slightly worse performance in conventional low-energy bins (\textasciitilde{} GeV). Further work is underway to improve the efficiency of the neural network at low energies.}

\FullConference{35th International Cosmic Ray Conference --- ICRC2017\\
		10--20 July, 2017\\
		Bexco, Busan, Korea}

\begin{document}
	\section{Introduction}
		\paragraph*{}The High-Altitude Water Cherenkov, HAWC, is a gamma-ray observatory that is able to detect primary particles in the energy range from 500 GeV to 100 TeV. HAWC is composed of 300 water Cherenkov detector (WCDs), each instrumented with 4 photomultiplier tubes (PMTs) that detect the Cherenkov light from charged particles in extensive air showers. Given that 99 \% of the events detected by HAWC come from cosmic-ray air showers, the study of gamma-ray sources requires the development of efficient techniques to achieve gamma-hadron separation \cite {2017arXiv170407411A, 2017ApJ...841..100A, 2017arXiv170101778A}.
		\paragraph*{}Currently,  HAWC uses two parameters (compactness ``$\mathcal{C}$'' and PINCness ``$\mathcal{P}$'') that help distinguish between gamma events and hadron events, this way of identifying particles is called standard  cut (for more details see the section 2.6 of \cite{2017arXiv170101778A}). Both parameters depend upon the spatial distribution of the charge deposited by the secondary particles of the shower inside of the array. All events are separated into bins ($\mathcal{B}$), that depend on the fraction of available PMTs, $f_{hit}$, that record light during during the event. The events satisfying certain optimized cut values on $\mathcal{P}$ and $\mathcal{C}$ are then considered photon candidates (Table \ref{Tab:bins} shows the optimal cut of each bin). In this work, we used neural networks to improve the gamma-hadron separation, in addition, we investigated the performance of the network when $\mathcal{C}$, $\mathcal{P}$ are used and when including new parameters in the inputs that could help distinguish between the two types of showers.
				\esqtab{The definition of the size bin $\mathcal{B}$ is given by the fraction of available PMTs, ($f_{hit}$), that record light during the event. Each bin has its own optimal cut on Compactnes ($\mathcal{C}$) and PINCness ($\mathcal{P}$). Events with a $\mathcal{P}$ less than indicated and a $\mathcal{C}$ greater than indicated are considered photon candidates \cite{2017arXiv170101778A}}{Tab:bins}{1}{1}{|c|c|c|c|}{	
				\hline
				$\mathcal{B}$ & $f_{hit}$ & $\mathcal{P}$ maximum  & $\mathcal{C}$ minimum\\
				\hline
				0 & 0.044 - 0.067 & <1.4 & >16 \\
				\hline
				1 & 0.067 - 0.102 & 2.2 & 7.0 \\
				\hline
				2 & 0.105 - 0.162 & 3.0 & 9.0\\
				\hline
				3 & 0.162 - 0.247 & 2.3 & 11.0\\
				\hline
				4 & 0.247 - 0.356 & 1.9 & 15.0\\
				\hline
				5 & 0.356 - 0.485 & 1.9 & 18.0\\
				\hline
				6 & 0.485 - 0.618 & 1.7 & 17.0\\
				\hline
				7 & 0.618 - 0.740 & 1.8 & 15.0\\
				\hline
				8 & 0.740 - 0.840 & 1.8 & 15.0\\
				\hline
				9 & 0.840 - 1.000 & 1.6 & 3.0\\
				\hline
			}
	\section{Training data set}\label{Sec:trainingset}
		\paragraph*{}The MC events were used to train the networks. Two types of particles were used: gamma and proton events, and they were simulated with a HAWC-250 configuration, that is, only 83 \%  of  the array was activated. We used protons to simulate cosmic rays because they constitute nearly 99 \% of the cosmic ray intensity. The conditions for selecting the events were:
			\begin{itemize}
				\item The event is well reconstructed, that is, the core and plane fit have been correctly computed. 
				\item The event falls within HAWC.
			\end{itemize}
		\paragraph*{}In addition, a uniform distribution of nHits\footnote{nHit is the number of PMTs activated during the event.} (an energy proxy) was chosen to better sample the high-energy range. Because if the natural distribution ($E^{-2.7}$) is used \cite{grieder2010extensive}, the network will have more experience to recognize events with low energy and have a poor performance at high energy.
	\subsection{Finding new parameters}\label{sec:newparameter}
		\paragraph*{} The performance of neural networks is heavily dependent on the input parameters used \cite{NNcourse}. Therefore, several parameters were analyzed. For the first attempt, the parameters that are in the HAWC data stream, were investigated. For the next attempt, new parameters were computed based on the timing information of the events. The following parameters are some of the ones that were evaluated for their G/H separation potential:
			\begin{itemize}
				\item Parameters that belong to the HAWC data stream
					\begin{itemize}
						\item planefitChi2:  The $\chi^2$ of a Gaussian plane fit to the shower front.
						\item \textbf{SFCFChi2}: The $\chi^2$ of a faster EAS core fit.
						\item \textbf{logmaxPE}: The PMT that has the maximum Photo-Electron (PE) charge during the event.
						\item logNPE: The logarithm of the total PE charge calculated in an event.
					\end{itemize}
				\item New time-based parameters
					\begin{itemize}
						\item deltaTime: It is the time difference between the trigger time of the first PMT and the last PMT that are triggered during the event
						\item TimeVariance: It is a simple measure of the scatter in PMT trigger time for a given event.
					\end{itemize}
			\end{itemize}
		\paragraph*{}The parameters with bold font (SCFCChi2 and logmaxPE) are used in this analysis because of their different profiles for gamma and protons, for example, the histogram of SFCFChi2 (Figure \ref{Fig:SFCFfeature}). The gamma histogram is skewed left,  whereas the hadron histogram has a more symmetric shape. Therefore they can be recognised with a cut.  We don't use the other parameters due to their histograms having the similar pattern cannot be separated with a cut like, such as deltaTime parameter (Figure \ref{Fig:dtime}).
			\onefig{0.7}{0.4}{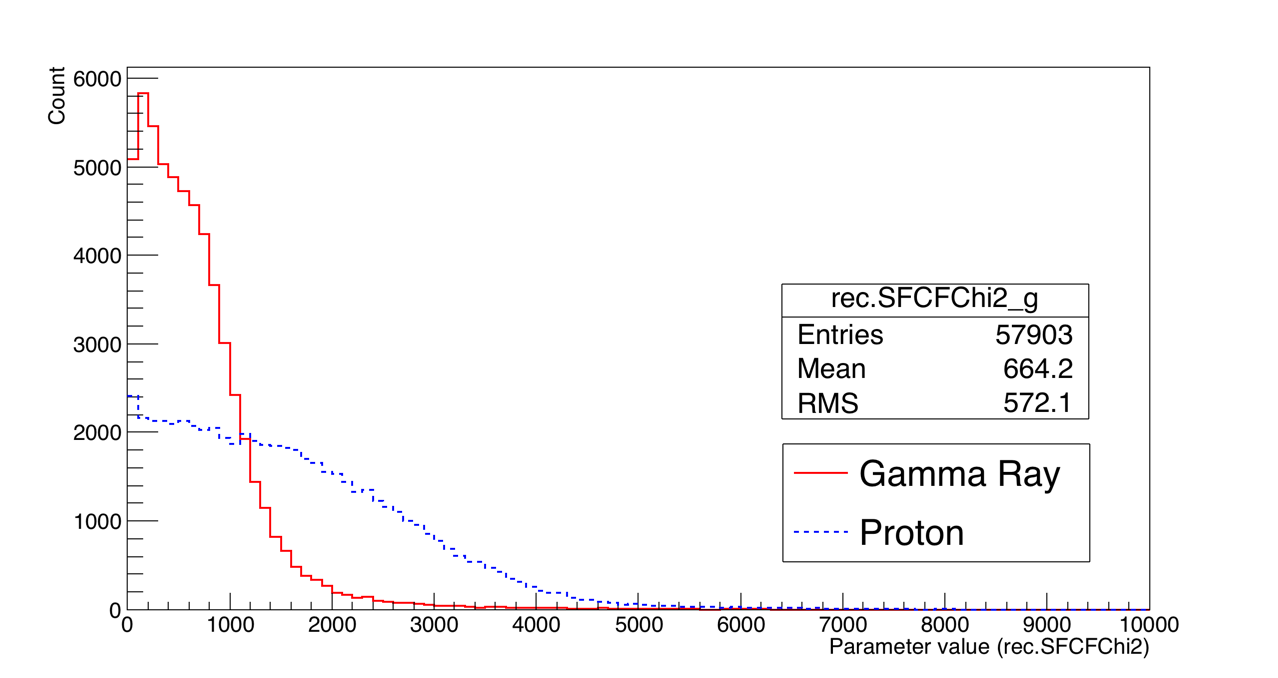}{The histograms of the $\chi^2$ of the core fit (SFCFChi2) using MC data. It is a potential parameter to use as input of the network for distinguishing between gammas and hadrons events since they have different shape and the histogram of gamma events is skewed right}{Fig:SFCFfeature}
			\onefig{0.7}{0.4}{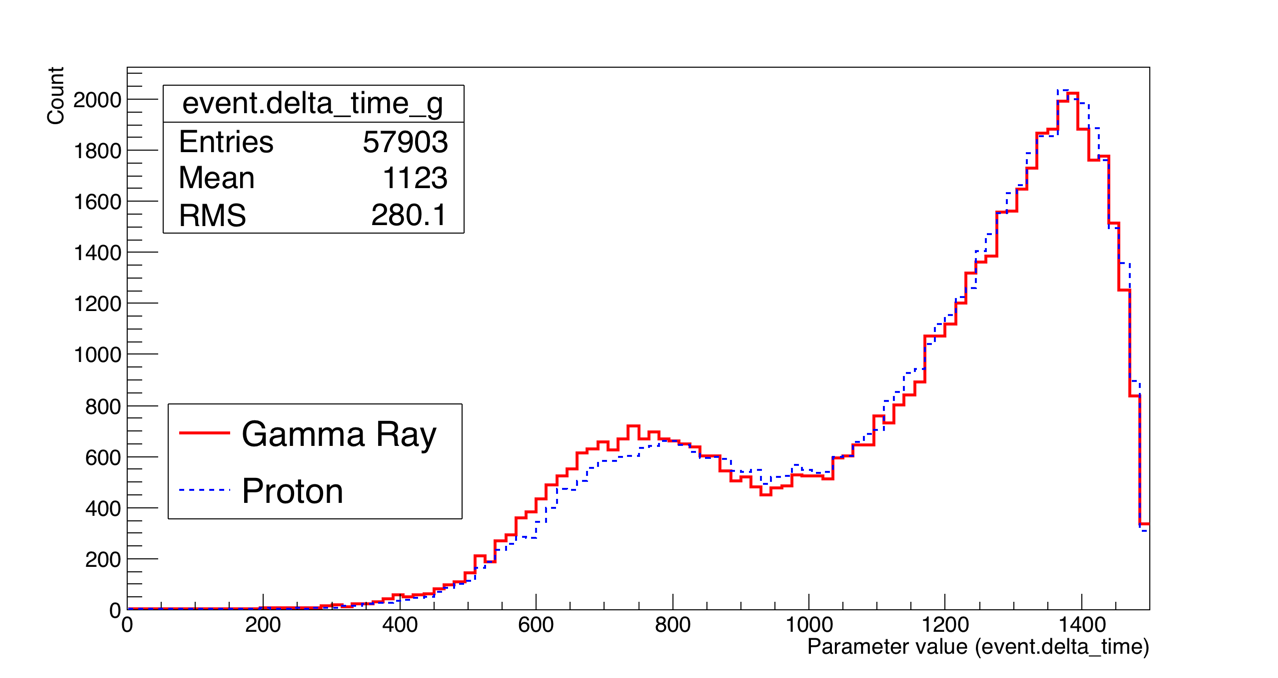}{The histograms of the deltatime for gamma and proton induced EAS using MC data, that is the time between first particle and last particle detected in the event. As both histograms have the same pattern (skewed left), this parameter is not useful as an input to the network}{Fig:dtime}
	\section{Training stage}
		\paragraph*{}In the training stage, HAWC simulations were used (see section \ref{Sec:trainingset}) for training the Neural Networks (NNs) to identify between primary particles. The NN used is a Multilayer Perceptron \cite{misc:pagrootmlp}. We defined a NN output value of 1 for gamma-like events, and 0 for hadron events: that is, the network must produce a value of 1 if the event is a gamma and 0 if it is a hadron (Figure \ref{Fig:nnoutput} shows an example of the output neural network). In order to identify the type of the events, a cut is used, in other words, if the output of the network is equal or greater than the cut value the event is considered like gamma event, in another case, it is a hadron event. The optimal cut is when it has an excellent hadron rejection but an acceptable gamma efficiency, it is when the Q factor has the maximum value, where is defined as $TPR~/~\sqrt {FPR}$, where $TPR$ is the fraction of correctly classified gamma events, also called gamma efficiency, and $FPR$ is the fraction of badly classified hadron events (hadron rejection).
			\doublefig{0.7}{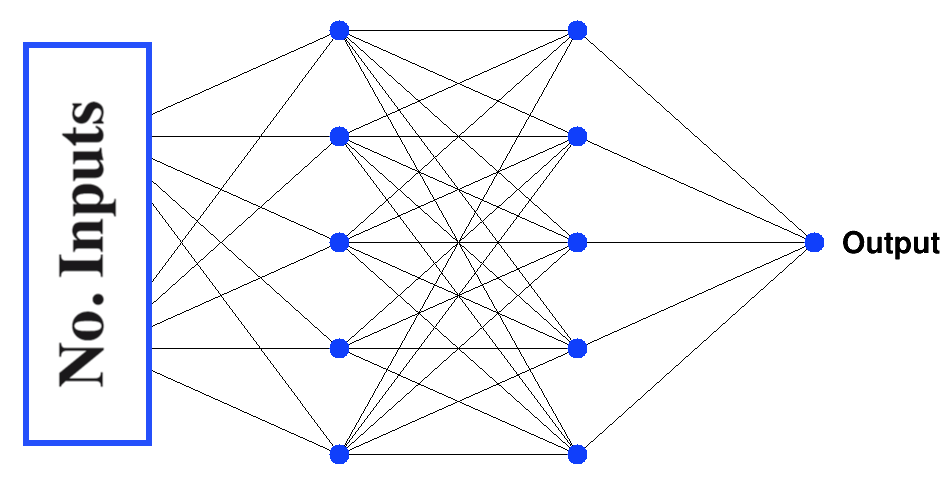} {Architecture of the neural network}{Fig:nnarchitecture}{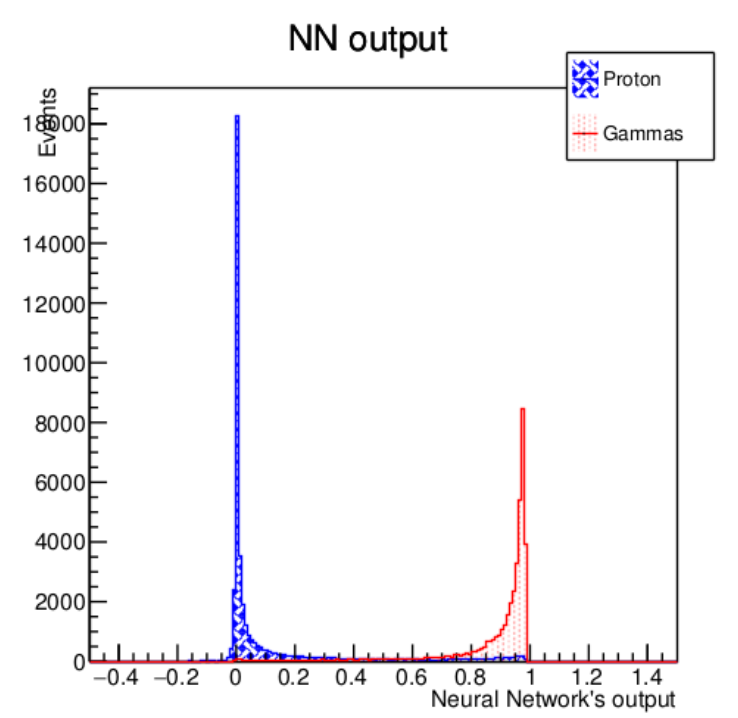}{Typical output of a neural network}{Fig:nnoutput}{In (a) shows the architecture of the network. The number of neurons in the first layer depends on the number of inputs that the network uses. In (b) is shown the histogram of the output of the NN2 after it was trained}{Fig:nn}
		\paragraph*{} Before adding the new parameters described in section \ref{sec:newparameter} to the network, one hundred training iterations of a simple 2-parameter network (Compactness and PINCness) were made, to see if the resulting optimal cut value for distinguishing between particles was sensitive to the random seed values given to the NN weights. The histogram on figure \ref{Fig:Optcutstable} shows the optimal cut values obtained from these trainings. The above figure shows that the training process appears to be giving stable cut values, which do not fluctuate too much. The slight variation of the maximum Q-factor (see figure \ref{Fig:qmaxstable}) can be attributed to the random nature of the initial NN weight seeding in the training process.		
			\doublefig{0.7}{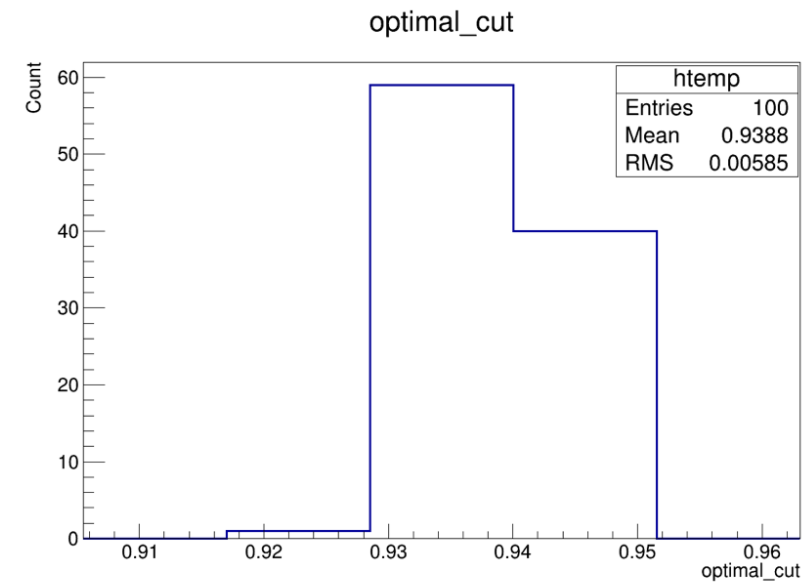}{Optimal cut}{Fig:Optcutstable}{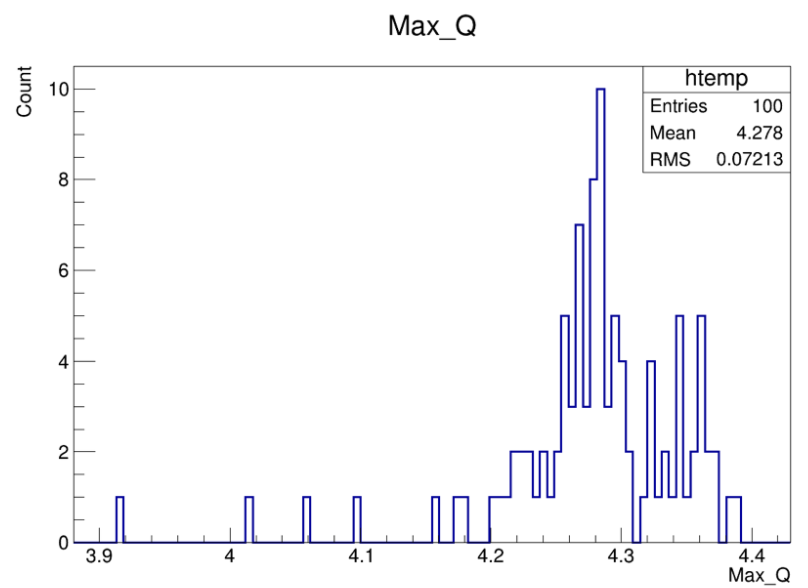} {Maximum Q factor}{Fig:qmaxstable}{One hundred NNs were trained in order to investigate the fluctation on the optimal cut and the Q factor in the training stage}{Fig:stable}
			\paragraph*{}For exploring the increase on G/H separation efficiency, three NNs were trained with similar architecture but different inputs (NoInputs:5:5:1, see Figure \ref{Fig:nnarchitecture}). The output layer defines the probability of being a gamma event or a hadron event. The networks that are used in this work are shown below:
			\begin{itemize}
				\item NN1: It uses three input parameters: $\mathcal{C}$, $\mathcal{P}$ and SFCFChi2 .
				\item NN2: It uses three input parameters: $\mathcal{C}$, $\mathcal{P}$ and logmaxPE.
				\item NN3: It uses four input parameters: $\mathcal{C}$, $\mathcal{P}$, SFCFChi2  and logmaxPE.
			\end{itemize}
	\section{Testing stage}
		\paragraph*{}The best way to compare the different performance between Standard Cut and the NNs is using HAWC data. We chose a set of well-reconstructed events\footnote{The core and plane are correctly computed.} within $\pm6^{\circ}$ of the Crab Nebula in the period from November 2014 to December 2015. One significance map per energy bin was made with a Simple Analysis Tools that is implemented in the official HAWC-software. From each map, the maximum significance was extracted, and given in the Table \ref{Tab:Crabdata}. The results in the bin 0 (low-energy)  show that the NN2 has better performances than the standard cut, suggesting that the logmaxPE parameter is useful for distinguishing events with low energy. Meanwhile, SFCFChi2 appears to improve the high-energy performance of the G/H separation process (bin 8 and 9),  with NN1 having the best results in this energy range. The combination of both parameters in the neural network, on the other hand, worsens the separation performances in almost every energy bin.
		\esqtab{The maximum significance in the Crab map obtained from each neural network trained in this work and from the standard cut}{Tab:Crabdata}{1}{1}{|c|c|c|c|c|}{	
			\hline
			$\mathcal{B}$ & Standard cut & NN1 & NN2 & NN3 \\
			\hline
			0 &  4.49 &  2.76 & 5.14 & 3.15 \\
			\hline
			1 &  9.60 &  3.92 & 8.24 & 2.67 \\
			\hline
			2 &15.39 & 6.89 & 14.28 & 5.15 \\
			\hline
			3 & 22.95 & 15.48 & 21.74 & 12.29 \\
			\hline
			4 & 30.13 & 29.75 & 30.98 & 25.44 \\
			\hline
			5 & 32.25 & 34.71 & 33.46 & 32.53 \\
			\hline
			6 & 32.75 & 34.77 & 31.88 & 31.74 \\
			\hline
			7 & 29.37 & 29.82 & 27.02 & 27.82 \\
			\hline
			8 & 28.73 & 31.25 & 28.04 & 28.89 \\
			\hline
			9 & 28.21 & 31.64 & 26.28 & 27.38\\
			\hline
		}
	\section{Conclusion}
		\paragraph*{} In this work, we trained 3 neural networks with $\mathcal{C}$, $\mathcal{P}$ and additional parameters, that had a potential for gamma-hadron separation in HAWC. In each network, a cut value on the NN's normalized output was selected by finding the maximal quality factor in the training phase of the analysis. Following the training on simulated shower events, the neural nets were tested using sample data from the Crab, and evaluating the maximum significance obtained on the source for all three types of NNs. The SFCFChi2 and logmaxPE parameters, which were selected for their higher gamma-hadron separation  potential, have allowed improve the high-energy and low-energy ranges of the data bins, respectivety. The combination of the two parameters inside the same neural network has, on the other hand, worse performances than the standard cuts, a behaviour which requires further investigations.
	\section*{Acknowledgments}
		\footnotesize{ We acknowledge the support from: the US National Science Foundation (NSF); the
US Department of Energy Office of High-Energy Physics; the Laboratory Directed
Research and Development (LDRD) program of Los Alamos National Laboratory;
Consejo Nacional de Ciencia y Tecnolog\'{\i}a (CONACyT), M{\'e}xico (grants
271051, 232656, 260378, 179588, 239762, 254964, 271737, 258865, 243290,
132197), Laboratorio Nacional HAWC de rayos gamma; L'OREAL Fellowship for
Women in Science 2014; Red HAWC, M{\'e}xico; DGAPA-UNAM (grants IG100317,
IN111315, IN111716-3, IA102715, 109916, IA102917); VIEP-BUAP; PIFI 2012, 2013,
PROFOCIE 2014, 2015;the University of Wisconsin Alumni Research Foundation;
the Institute of Geophysics, Planetary Physics, and Signatures at Los Alamos
National Laboratory; Polish Science Centre grant DEC-2014/13/B/ST9/945;
Coordinaci{\'o}n de la Investigaci{\'o}n Cient\'{\i}fica de la Universidad
Michoacana. Thanks to Luciano D\'{\i}az and Eduardo Murrieta for technical support.
		}
	\bibliographystyle{JHEP}
	\bibliography{biblio}

\providecommand{\href}[2]{#2}\begingroup\raggedright\begin{thebibliography}{1}

\bibitem{2017arXiv170407411A}
A.~U. Abeysekara et~al., {\it {The HAWC real-time flare monitor for rapid
  detection of transient events}},  {\em ArXiv e-prints} (Apr., 2017)
  [\href{http://arxiv.org/abs/1704.0741}{{\tt arXiv:1704.0741}}].

\bibitem{2017ApJ...841..100A}
A.~U. Abeysekara et~al., {\it {Daily Monitoring of TeV Gamma-Ray Emission from
  Mrk 421, Mrk 501, and the Crab Nebula with HAWC}},  {\em The Astrophysical
  Journal} {\bf 841} (June, 2017) 100,
  [\href{http://arxiv.org/abs/1703.0696}{{\tt arXiv:1703.0696}}].

\bibitem{2017arXiv170101778A}
A.~U. Abeysekara et~al., {\it {Observation of the Crab Nebula with the HAWC
  Gamma-Ray Observatory}},  {\em ArXiv e-prints} (Jan., 2017)
  [\href{http://arxiv.org/abs/1701.0177}{{\tt arXiv:1701.0177}}].

\bibitem{grieder2010extensive}
P.~Grieder, {\em Extensive Air Showers: High Energy Phenomena and Astrophysical
  Aspects - A Tutorial, Reference Manual and Data Book}.
\newblock Astrophysics and space science library. Springer Berlin Heidelberg,
  2010.

\bibitem{NNcourse}
{Geoffrey Hinton}, ``{Neural Networks for Machine Learning | Coursera}.''
  \url{https://www.coursera.org/learn/neural-networks}.
\newblock Accessed January, 2016.

\bibitem{misc:pagrootmlp}
{Christophe Delaere}, ``{class TMultiLayerPerceptron}.''
  \url{http://root.cern.ch/root/html/TMultiLayerPerceptron.html}.
\newblock Accessed January, 2014.

\end{thebibliography}\endgroup
\end{document}